\documentclass[prb,twocolumn,tbtags,superscriptaddress,floatfix]{revtex4}
\usepackage{amssymb}
\usepackage{graphicx,amsmath}
\usepackage{bm}
\usepackage{times}
\usepackage{ulem}

\begin{document}

\title{Landau-Zener-St\"{u}ckelberg-Majorana lasing in circuit QED}

\begin{abstract}
We demonstrate amplification (and attenuation) of a probe signal by a driven two-level quantum system in the Landau-Zener-St\"{u}ckelberg-Majorana regime by means of an experiment, in which a superconducting qubit was strongly coupled to a microwave cavity, in a conventional arrangement of circuit quantum electrodynamics. Two different types of flux qubit, specifically a conventional Josephson junctions qubit and a phase-slip qubit, show similar results, namely, lasing at the working points where amplification takes place. The experimental data are explained by the interaction of the probe signal with Rabi-like oscillations. The latter are created by constructive interference of Landau-Zener-St\"{u}ckelberg-Majorana (LZSM) transitions during the driving period of the qubit. A detailed description of the occurrence of these oscillations and a comparison of obtained data with both analytic and numerical calculations are given.
\end{abstract}

\date{\today }
\author{P.~Neilinger}
\affiliation{Department of Experimental Physics, Comenius University, SK-84248
Bratislava, Slovak Republic}
\author{S.~N. Shevchenko}
\affiliation{B. Verkin Institute for Low Temperature Physics and Engineering, 61103 Kharkov, Ukraine}
\affiliation{V. Karazin Kharkov National University, 61022 Kharkov, Ukraine}
\author{J.~Bog\'{a}r}
\affiliation{Department of Experimental Physics, Comenius University, SK-84248
Bratislava, Slovakia}
\author{M.~Reh\'{a}k}
\affiliation{Department of Experimental Physics, Comenius University, SK-84248
	Bratislava, Slovakia}
\author{G. Oelsner}
\affiliation{Leibniz Institute of Photonic Technology, D-07702 Jena, Germany}
\author{D.~S.~Karpov}
\affiliation{B. Verkin Institute for Low Temperature Physics and Engineering, Kharkov,
Ukraine}
\author{U.~H\"{u}bner}
\affiliation{Leibniz Institute of Photonic Technology, D-07702 Jena, Germany}
\author{O.~Astafiev}
\affiliation{Physics Department, Royal Holloway, University of London, Egham, Surrey TW20
0EX, United Kingdom}
\affiliation{National Physical Laboratory, Teddington, TW11 0LW, United Kindom}
\affiliation{Moscow Institute of Physics and Technology, Dolgoprudny, 141700, Russia}
\author{M.~Grajcar}
\affiliation{Department of Experimental Physics, Comenius University, SK-84248
Bratislava, Slovakia}
\affiliation{Institute of Physics of Slovak Academy of Sciences, D\'{u}bravsk\'{a} cesta,
Bratislava, Slovak Republic}
\author{E.~Il'ichev}
\affiliation{Leibniz Institute of Photonic Technology, D-07702 Jena, Germany}
\affiliation{Novosibirsk State Technical University, 630092 Novosibirsk, Russia}
\maketitle

\section{Introduction}\label{Introduction}

Although the Landau-Zener (LZ) problem was extensively studied already in the 30s of the last century \cite{Landau32a,Landau32b,Zener32}, nowadays, new phenomena are revealed as a result of dissipation \cite{Ao89}, environmental noise \cite{Blattmann15}, as well as measurement back-action \cite{Haikka14} on the LZSM interference. It has been shown that interferometry can be very useful in resolving of both spectroscopic \cite{Berns08} and dissipative environmental \cite{Forster13} information about an investigated system. Although LZSM increases the occupation probability of the excited state, population inversion cannot be achieved for an isolated two-level system without relaxation, and coupling to a measurement device (and/or the environment) should depopulate the excited state even more. Fortunately, what at first seems counter-intuitive, a "continuous measurement" of the two-level system by a detector or an "environment" can lead to a significant excitation in spite of the decay.\cite{Haikka14}

In this paper, we report on the experimental observation of LZSM interference patterns through the amplification/attenuation of a probe signal (stimulated emission/absorption) as well as lasing (free emission) in a driven two-level quantum system coupled to a microwave resonator \cite{Ilichev03,Wal04,Ble04} under an external off-resonant drive. The observed interference patterns are studied by the analytic approach of the so-called adiabatic-impulse method (AIM), Ref.~\onlinecite{Shevchenko10} and references therein. The AIM was shown to describe well quantitatively the  dynamics of the two-level quantum system in a broad parameter range.\cite{Ashhab07, Ferron10, Shevchenko12, Zhou14, Silveri15} This method, which essentially describes the evolution of a system as the alteration of adiabatic stages of evolution with stroboscopic non-adiabatic transitions, the LZSM transitions \cite{DiGiacomo05}, was recently studied for a number of quantum systems driven by different periodical fields.\cite{Oliver05, Sillanpaa06, Wang10, Forster13, Silveri15} In particular, it was predicted that interference between multiple LZSM transitions can produce periodic oscillations of the level occupations. Quite recently these oscillations have been observed in the time domain for a spin ensemble by making use of NV centres in diamond \cite{Zhou14}. Since they are reminiscent of Rabi oscillations, they can be termed as LZSM-Rabi-like oscillations, however, for brevity, we will call them Rabi-like oscillations.
Oscillations of the level occupation in resonantly driven two-level quantum systems are the core of different spectroscopic techniques. One interesting aspect, which was extensively studied recently, is the amplification/attenuation of microwave quantum signals.\cite{Hauss08, Oelsner13, Koshino13, Liu14, Neilinger15, Karpov2016}. The Rabi oscillations are adjusted by driving to match the weak (probe) signal frequency, $\Omega _{\mathrm{R}}\approx \omega _{\mathrm{p}}$. Then, the resonant interaction between the two-level quantum system and the probe signal results in energy exchange between these two subsystems. Thus, it is quite natural, similar to the use of Rabi oscillations, to exploit the Rabi-like oscillations for the processing of microwave quantum signals. Moreover, this approach can account for multiple interactions in a single calculation and thus can be simply used in parameter regions where it would be necessary for the rotating wave approximations with different frequencies to be applied at once.\cite{Shevchenko14}
This qualitative analysis, which provides the observed contours of the LZSM interference patterns, is corroborated by numerical simulations of a multi-level qubit-resonator system based on the adiabatic-impulse model.

This paper is arranged as follows. In Sec.~\ref{Experiments}, we present our experimental results obtained in two experiments carried out on two different types of superconducting flux qubits. In Sec.~\ref{Theory}, we analyze the oscillations of the upper-level occupation probability of a driven two-level system and describe the interaction of a driven two-level system and a resonator in terms of Rabi-like oscillations. Two regimes, depending on the ratio of the drive frequency $\omega$ and the minimal splitting of the two-level system $\Delta$ relevant to our experiment are distinguished, namely, the slow-passage limit ($\Delta /\protect\hbar\omega \gg 1 $) and the fast-passage limit ($\Delta /\protect\hbar\omega \ll 1 $). In Sec.~\ref{Numerical model}, a numerical computation of the average photon number in the resonator is carried out on a driven two-level system strongly coupled  to a single-mode radiation field of a quantized resonator, creating a multi-level qubit-resonator system.  The simulation reveals LZSM interference patterns in the average photon number which are in good agreement with the one obtained by the analytical approach of the Rabi-like oscillations. In Appendices A and B we provide additional  details on the theory of Rabi-like oscillations and the experimental set-up, respectively.

%Recently, the resonant case, where the Rabi oscillations are adjusted by driving to match a weak (probe) signal frequency, was extensively studied \cite{Hauss08, Oelsner13, Koshino13, Liu14, Neilinger15}. Similar to the presented experiments, amplification/attenuation of the microwave quantum signals was observed. Thus, it is quite natural, similar to the use of Rabi oscillations, to exploit Rabi-like oscillations for processing of microwave quantum signals. 
%The LZSM interference is described in terms of "Rabi-like" oscillations, reminiscent of Rabi oscillations in a strongly driven two-level system. 
\section{Experiments}\label{Experiments}

Our experiments were carried out on two different types of superconducting flux qubits. They are the flux qubit based on conventional Josephson junctions \cite{Mooij99}, and the phase-slip (QPS) qubit, a novel qubit type, based on nanowires made from thin films of niobium nitride (NbN).\cite{Astafiev12} 

The aluminium Josephson junction flux qubit is coupled to a niobium resonator with resonance frequency $\protect\omega _{\mathrm{r}}/2\protect\pi \approx 2.481$~GHz, and quality factor $Q \approx 9\;000$ for the fundamental half-wavelength mode. The resonator is in the overcoupled regime, thus the measured loaded quality factor is governed by its external quality factor.\cite{Goppl2008} The Josephson junction qubit tunneling energy is $\Delta/h \approx $12.2 GHz and represents the minimal level splitting of the qubit states. The energy bias of the qubit depends on the external magnetic flux $\Phi$ as $\varepsilon_0=2I_{p}(\Phi-\Phi_0/2)$, where $\Phi_0$ is the magnetic flux quantum and $I_{p}$ is the persistent current of the qubit. The latter takes a value of $I_p \approx $ 35 nA for the conventional flux qubit.\cite{Oelsner10} The coherence time of the qubit $T_2 \approx$ 100 ns and the qubit-resonator coupling $g\approx 70$~MHz were estimated from a fit of the resonator transmission at $\protect\omega_{r}$ as a function of energy bias taking into acount multiphoton processes.\cite{Omelyanchouk10,Niemczyk2010,Chen2016} Other details on the experiment can be found in Appendix B.
 
The quantum phase slip qubit is a several microns sized loop, patterned from a thin (about 2 nm ) film of NbN.\cite{Peltonen13} The persistent current of the qubit is $I_p$ = 30 nA and the tunneling energy $\Delta/h$ = 6.12 GHz. The resonator fundamental frequency is $\omega_r/2\protect\pi \approx$ 2.3 GHz, however, the measurements presented here are done at the third mode at $\omega_{3}/2\pi$ = 6.967 GHz, where the quality factor is $Q \approx$ 500. The coupling strength between the qubit and the resonator is of the order of 100 MHz and the qubit coherence time is $T_2 \approx$ 25 ns.

\begin{figure}[]
	\includegraphics[width=7.5cm]{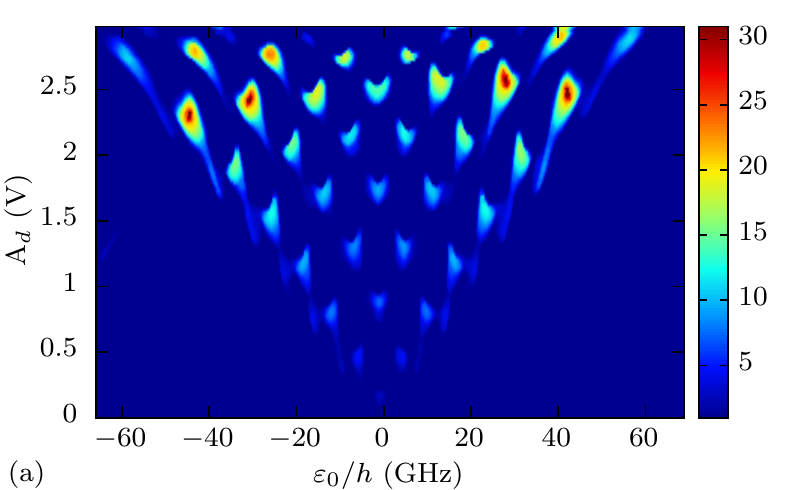}
	\includegraphics[width=7.5cm]{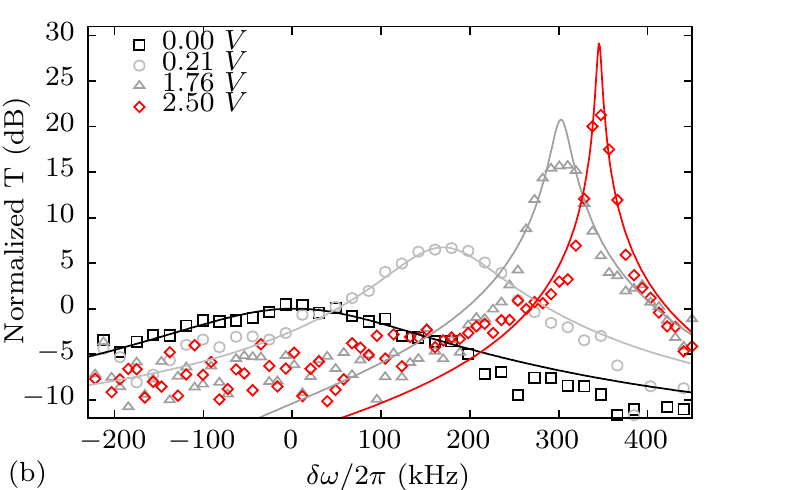}
	\includegraphics[width=7.5cm]{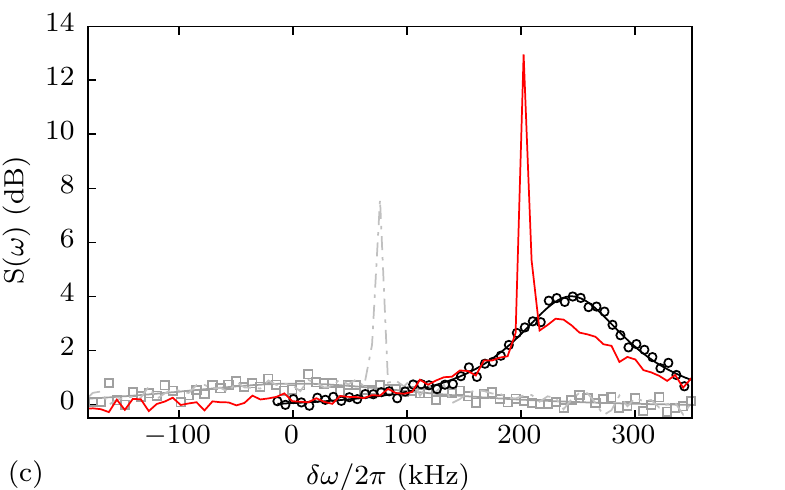}		
	\caption{(color online). (a) Normalized power transmission maximum in dB for the resonator coupled to the Josephson junction qubit as a function of the qubit energy bias and the driving amplitude at frequency $\protect\omega/2\pi =7.444$~GHz. The transmission maxima were obtained by a Lorentzian fit of the measured transmission spectra. The transmission is quasi-periodically increased and suppressed, revealing characteristic LZSM interference patterns. Here: $\Delta/h=12.2$~GHz, $\protect\omega _{\mathrm{r}}/2\protect\pi =2.481$~GHz and the ratio $\Delta /\hbar \protect\omega \approx 1.6$, which is closer to the slow-passage limit (see main text). 
	(b) Normalized power transmission spectra (data points) at driving amplitudes corresponding to the transmission maximums at zero bias and the corresponding Lorentzian fits (solid lines). Here $\delta\protect\omega$ is the detuning of the weak probe signal from the resonator's fundamental frequency $\delta\protect\omega = \protect\omega _{\mathrm{p}}-\protect\omega _{r}$.
		(c) Spectral power density of the microwave radiation emitted by the resonator. The data points correspond to emission without driving (squares) and driving amplitude set to 2.40 V (circles) at zero bias. The black dashed lines corresponds Lorentzian fits. Similar to the transmission measurements, for driving turned ON, the resonator emission is increased and it's bandwidth is narrowed. To illustrate the amplification, a weak probe signal is applied in the bandwidth of the resonance in the absence of driving (gray dashed line) and with driving (red solid line). For driving set ON the emission is locked to the probe frequency and energy is transferred, which is visible by the shrinkage of the Lorentzian shaped emission curve.}
	\label{Fig:Pavol_s}
\end{figure}

\begin{figure}[]
	\includegraphics[width=7.5cm]{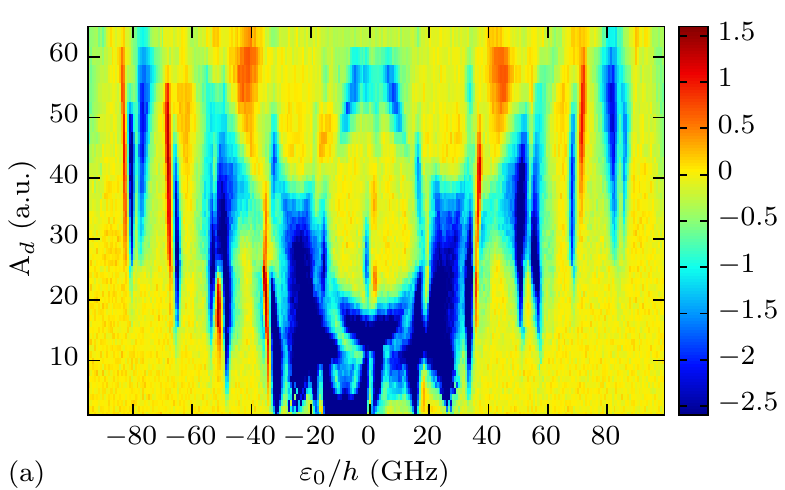}
	\includegraphics[width=7.5cm]{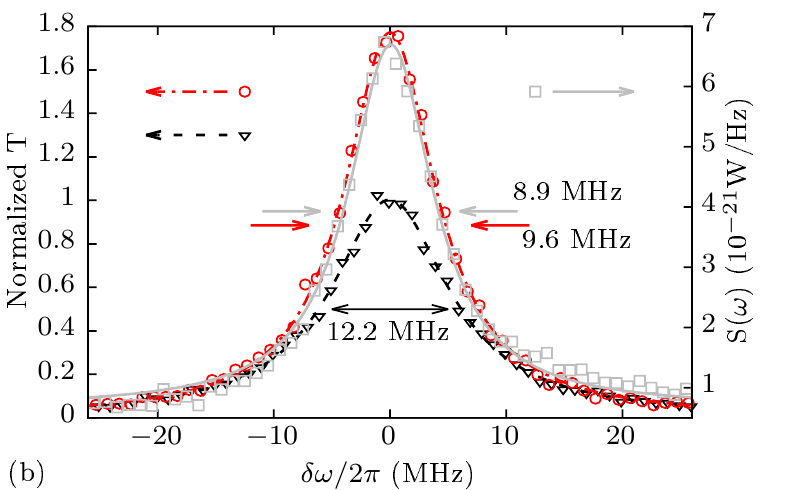}
	\caption{(color online). (a) Spectral power density in dB emitted by the resonator with the QPS qubit. LZSM lasing for small ratio $\Delta /\protect\hbar\omega =$ 0.43, which corresponds to the fast-passage limit (see main text). The position of the
		resonant amplification and attenuation points corresponds to the one- and
		two- photon Rabi oscillations, $\Omega _{\mathrm{R}}^{(k)}=\protect\omega _{%
			\mathrm{d}}$ with $k=1,2$. Here, $\Delta /h= 6.12$ GHz, $\protect\omega = 2\pi\times16.3$ GHz and the emission is measured  at $\omega_3/2\pi$ = 6.967 GHz.
			(b) Normalized power transmission through the resonator at $\protect\omega_{3}$ without (black triangles) and with driving at 16.3 GHz (red open dots) and the measured emission under the same conditions with driving (gray squares). The lines correspond to Lorentzian fits.}		
		\label{Fig:Oleg_s}
	\end{figure}
	
Amplification of the traversing signal through the resonator, as well as free emission from the resonator with the qubits are studied under an external off-resonant drive in the LSZM regime. The experimental results are presented separately for the system with the Josephson junction qubit and the QPS qubit in Fig.~\ref{Fig:Pavol_s} and Fig.~\ref{Fig:Oleg_s}, respectively. For both systems, transmission measurements (carried out by a vector network analyzer) and emission measurements (carried out by a power spectrum analyzer) are compared. 

The power transmission spectrum of the resonator T($\omega$) coupled to the Josephson junction qubit, measured by a weak probing signal $\protect\omega_{p}$ close to the resonator fundamental frequency $\protect\omega_{r}/2\pi \approx 2.481$~GHz, was characterized as a function of the drive amplitude $A_d$ at frequency  $\protect\omega/2\pi =7.444$~GHz and the dc bias of the qubit $\varepsilon_0$. The transmission spectrum at each working point $\left[\varepsilon_0,A_d \right]$ was fitted to Lorentz function to estimate the power transmission maximum, resonance frequency and the quality factor of the resonator. All of these parameters strongly depend on the driving amplitude and the qubit bias. The normalized power transmission maximum, plotted as a colormap in Fig.~\ref{Fig:Pavol_s}(a), reveals characteristic LZSM interference patterns with quasi-periodic maxima and minima. This increase of the power transmission is accompanied by significant bandwidth narrowing of the resonance curve and a slight shift of the resonance frequency. The measured normalized power transmission spectra for driving amplitudes corresponding to the transmission maxima at zero bias are shown in Fig.~\ref{Fig:Pavol_s}(b). By increasing the driving amplitude from zero to 0.21 V, 1.76 V, and 2.50 V the maximal transmission of the resonator increases and the bandwidth subsequently decreases from 280 kHz to 121 kHz, 20 kHz, and 5.4 kHz. These values are obtained from the fit of experimental data (points) to Lorentz functions (solid lines). 

To show that both emission and transmission measurements reveal the same phenomena, namely the amplification and suppression of electromagnetic waves passing the resonator, we study the spectral power density spectra of the microwave radiation emitted by the resonator under driving. In Fig.~\ref{Fig:Pavol_s}(c) the resonator emission at zero bias and driving turned OFF (squares) and turned ON (amplitude set to p = 2.40 V, circles) are shown. 
For driving ON, the emission is increased, the bandwidth narrows from 285 kHz to 100 kHz and resonance shifts by 245 kHz. These parameters were obtained from a Lorentzian fits (black dashed lines).
Further, to illustrate the amplification observed by the transmission of the resonator, a weak probe signal in the bandwidth of the resonator was applied for both cases - the driving turned on and off (gray dashed line and red solid line). In the absence of driving, the probing signal is visible as a narrow peak in the power spectral density added to the wide Lorentzian background (grey dashed curve). For driving ON, the probe signal is amplified  and the emission is locked at frequency $\omega_p$, visible as energy transfer (the area between dashed and dash-dotted red line in Fig.~\ref{Fig:Pavol_s}(b)) to the peak at $\omega_p$. This effect of injection-locking was already observed for single artificial-atom lasing in Ref.~\onlinecite{Astafiev07}.

Similarly, the QPS qubit was studied for amplification (by VNA) and emission (PSA). Fig.~\ref{Fig:Oleg_s}(a) demonstrates power transmission versus bias $\varepsilon_{0}$ and driving. The emission is measured at $\omega_3/2\pi$ = 6.967 GHz, while the driving frequency is $\omega/2\pi$ = 16.3 GHz. Although the pattern is different, it essentially demonstrates the same behavior. We observe absorption (blue areas) and emission (vertical red stripes) corresponding to different multi-photon processes. Fig.~\protect\ref{Fig:Oleg_s}(b) demonstrates the square amplitude of transmission through the resonator at $\omega_3$ (black triangles), which is 12.2 MHz at the full width at half maximum (FWHM) without driving, determined by the photon decay rate. When the driving at 16.3 GHz is ON, the transmitted signal is amplified (red open dots) by a factor of two in power and the FWHM becomes narrower, reaching 9.6 MHz. The measured emission under the same conditions shows a high and narrow peak of 8.9 MHz width (gray crosses), which corresponds to roughly 100 photons in the resonator.
		
The observed experimental results clearly demonstrate amplification of the transmitted signal with certain indication of a lasing effect at the LZSM interference maxima, since a narrowing of the bandwidth and injection-locking were convincingly detected. 

\section{Rabi-like oscillations}\label{Theory}

\begin{figure}[tph]
	\includegraphics[width=4cm]{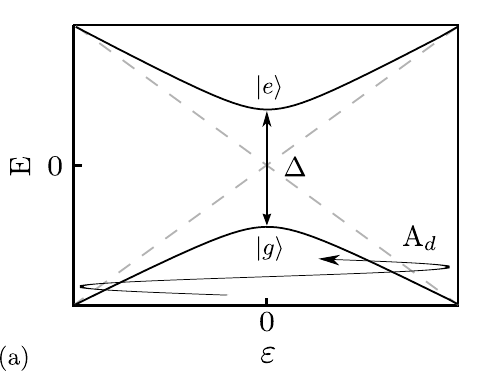}
	\includegraphics[width=4cm]{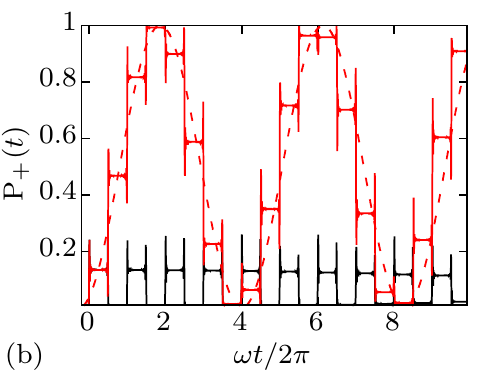}
	\includegraphics[width=4cm]{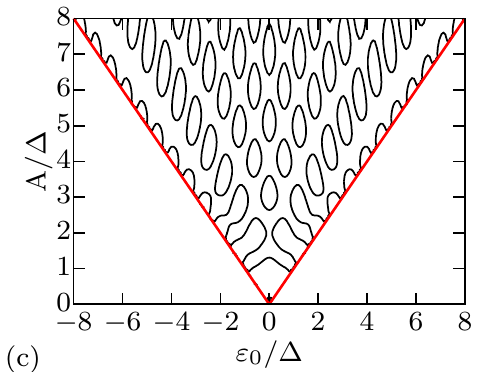}
	\includegraphics[width=4cm]{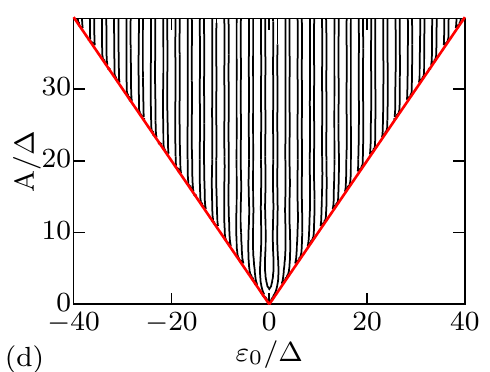}
	\caption{(color online). (a) Energy levels $E$ of a  two-level system as a function of the energy bias $\protect\varepsilon $ of a superconducting qubit with energy level splitting $\Delta $. 
The energy bias $\protect\varepsilon $ is driven with a sinusoidal driving signal at frequency $\protect\omega$. Under driving, the two-level system undergoes subsequent LZSM transitions. (b) Crossover between the subsequent LZ transitions and Rabi-like oscillation, resulting from constructive interference. The upper-level occupation probability is plotted as a function of time for many periods of the driving field. For $\protect\varepsilon _{0}=0$, $A/\Delta = 15.71$, and $\hbar \protect\omega /\Delta \approx 0.05$, which corresponds to $P_{\mathrm{LZ}}=0.13\ll 1$ the time evolution shows destructive interference of subsequent LZ transitions (black curve). If the amplitude is slightly varied to $A/\Delta =15.75$ the constructive interference leads to Rabi-like oscillations approximated by the dashed sinusoidal line. The frequency of the Rabi-like oscillations is given by $\Omega \ll \protect\omega $, see Eq.~(\protect\ref{cos(fi)}). Note that these Rabi-like oscillations appear far from resonance, at $\protect\omega \ll \Delta E/\hbar $. Position of the expected resonant interactions between the driven qubit and the weak probe signal, as defined by Eq.~(\protect\ref{condition}), are shown for the slow-passage limit in (c) and the fast-passage limit in (d). The following parameters were taken: $\protect\omega _{\mathrm{p}}/2\protect\pi =2.5$ GHz, $\protect\omega =3\protect\omega _{\mathrm{p}}$, $\Delta /h=12.2$ GHz $>\protect\omega /2\protect\pi $ for (c) and $\Delta /h=3$GHz $\ll \protect\omega /2\protect\pi $ for (d). The inclined red lines in (c) and (d) mark the region of the validity of the theory: $\protect\varepsilon _{0}<A$, which means that the system experiences avoided level crossings.} 
		\label{Fig:P(t)}
	\label{Fig:En_lev}
	\end{figure}

In theory, a driven tunable two-level system can be described using Pauli matrices $\sigma _{x,z}$ by the Hamiltonian $H_q(t)=-\frac{1}{2}\left( \Delta \sigma _{x}+\varepsilon (t)\sigma _{z}\right)$, with the constant term $\Delta $ (tunneling energy) and the time-dependent one $\varepsilon(t)=\varepsilon _{0}+A\sin \omega t$, where $A$  is the bias amplitude of the field applied at frequency $\omega$. The respective Schr\"{o}dinger equation can not be solved analytically in general case, and thus a variety of theoretical tools are applied to this "simplest non-simple quantum problem". \cite{Berry95} Arguably, the most intuitive tool is the adiabatic-impulse method (AIM); see Ref.~\cite{Shevchenko10} and references therein. We consider here the adiabatic limit, where the frequency $\omega$ is a small parameter ($\hbar \omega < \Delta, A$).
When driven, the system follows its eigenstates $|g\rangle$ and $|e\rangle$, for ground and excited states, respectively. The corresponding eigenenergies of the Hamiltonian $H_q$ are $E_{g,e}(t)=\mp \frac{1}{2} \sqrt{\Delta ^{2}+\varepsilon^{2}}$. The energy levels are depicted in Fig.~\ref{Fig:En_lev}(a). Close to the degeneracy point, when $\varepsilon(t) = 0$, tunneling between the two states is possible. Note, that during one period of driving, this point is reached two times, denoted as t$_{1}$ and t$_{2}$. The probability of tunneling between the states is $P_{\mathrm{LZ}}=\exp \left(-2\pi \delta \right)$, known as the Landau-Zener (LZ) probability, where $\delta =\Delta^{2}/4\hbar \omega \sqrt{A^{2}-\varepsilon _{0}^{2}}$ is the adiabaticity parameter. One can distinguish two extreme regimes: (i) the slow-passage limit ($\delta >1$ such that $P_{\mathrm{LZ}}\ll 1$) and (ii) the fast-passage limit ($\delta \ll 1$ such that $1-P_{\mathrm{LZ}}\ll 1$).

During one period of the drive, the wave function accumulates the phases $\widetilde{\zeta }_{1,2}=\frac{1}{2\hbar }\int_{1,2} \sqrt{\Delta^{2}+\varepsilon (t)^{2}}dt+\widetilde {\varphi }_{\mathrm{S}}$, where the first dynamical part is defined by the adiabatic evolution and the index denotes the integration intervals between the LZ transitions $(t_{1},t_{2})$ and $(t_{2},t_{1}~+~2\pi /\omega )$. The second part is acquired during the LZ transition and it is depending on the adiabacity parameter  $\delta$ as $\widetilde{\varphi }_{\mathrm{S}}=-\frac{\pi}{4}+\delta (\ln \delta -1)+\arg \Gamma (1-i\delta )$, with $\Gamma$ being the gamma function and $\arg$ denotes the argument of a complex number. 
Numerically, the probability amplitudes from the Schr\"odinger equation may be found, as demonstrated in Appendix A. 
They are plotted in Fig.~\ref{Fig:En_lev}(b). Note, that AIM predicts a step-like evolution. In the case of constructive interference, during many driving periods, the upper level occupation probability increases, up to a maximal value of $P_+ = 1$. In the long run, this displays an almost periodic behavior, with slow oscillations reminiscent of the Rabi oscillations, which we will call Rabi-like oscillations.
In the general case (see Appendix A), the AIM allows an analytical solution for the frequency of these Rabi-like oscillations, which is given by
\begin{equation}
\Omega =\frac{\omega }{\pi }\arccos \left\vert \left( 1-P_{\mathrm{LZ}%
}\right) \cos \zeta _{+}-P_{\mathrm{LZ}}\cos \zeta _{-}\right\vert.
\label{cos(fi)}
\end{equation}%
In our consideration, the most interesting case is when these driven
(slow) oscillations come in resonance with our probe signal:%
\begin{equation}
\Omega (\varepsilon _{0},A)=\omega _{\mathrm{p}}  \label{condition}
\end{equation}
providing energy exchange between the qubit and the resonator.

With Eq.~(\ref{condition}), the position of expected resonances between the Rabi-like oscillations and the resonator mode can be predicted for a qubit coupled to a quantized resonator field, plotted in Fig.~\ref{Fig:P(t)} for the slow and fast-passage limits.

The shape of the interference fringes (see Fig.~\ref{Fig:P(t)}) qualitatively corresponds to the measured results for the standard qubit (Fig.~\ref{Fig:Pavol_s}) and for the QPS qubit (Fig.~\ref{Fig:Oleg_s}). As we found for our samples, they work in the slow-passage and in the fast-passage limit, respectively. Note, this is only given by the relation of the energy gap and driving frequency, and it is not a unique feature of the chosen qubit types.	

The LZSM theory, which does not include relaxation and dephasing, does not provide population inversion. However, certain analogy between driven systems exhibiting Rabi oscillations ("resonant" case) and the Rabi-like oscillations ("off resonant" case) can be demonstrated. Similar to the resonant case, when the system's energy levels are coupled by resonant interaction (usually with small detuning $\delta = \omega - \omega_q << \omega$), the levels are coupled via LZ transitions, providing level splitting proportional to the frequency of the Rabi-like oscillations. This means that the energy level structure is very similar for both cases. 

In order to analyze the amplification and damping by making use of the interaction picture, the expression for the coupling between the resonator and the flux qubit $H_c = M I_q I_r$ (where $M$ is the mutual inductance between them and $I_q$ and $I_r$ the respective currents in the qubit and the resonator) should be transformed to $H_c = M I_p I_{r0} \sigma_z \left( a e^{-i\omega_p t} + a^\dagger e^{i\omega_p t} \right) $. Here, $I_{r0}$ is the zero point current amplitude of the resonator. For both Rabi and Rabi-like oscillations, the periodic change of the population of the states is expressed as $\sigma_z e^{i \Omega t}$. If $\Omega=\omega_p$, depending on the sign of the Rabi or Rabi-like frequency, a time average of $H_c$ will define whether photons are created ($a^\dagger$) in the cavity or absorbed ($a$) from the cavity. A possible sign change is expected, when at a working point, the ground state with $N$ photons lays above the excited state with $N-1$ photons. The detailed role of relaxation in determining the sign of the detuning and the amplitude of the oscillations requires further analysis.

\section{Numerical model}\label{Numerical model}

%We demonstrate the applicability of the AIM method by simulating the time evolution of a strongly driven qubit coupled to the quantized photon field of a resonator and calculating the average photon number $\bar{n}$ in the resonator as a function of the qubit energy bias and the driving amplitude in the absence of a probing signal. We consider this system to be described by the multiphoton Jayness-Cummings Hamiltonian \cite{Vogel2006}

%%%%%%%%%%%%%%%%%%%%%%%%%%%
In this section, we introduce a multi-level model of a two-level system  strongly coupled to a single-mode radiation field of a quantized resonator and numerically simulate the time evolution of the photon number occupancy in the resonator under off-resonant drive. The coupled qubit resonator system can be described by the multi-photon Jaynes-Cummings model with Hamiltonian (see, for example, Ref.~\onlinecite{Vogel2006}):   

\begin{equation}
H=\frac{\hbar \omega _{q}}{2}\sigma _{z}+\hbar \omega _{r}\left(a^{\dag }a+\frac{1}{2}\right) +\hbar g_{(k)}(a^{\dag (k)}\sigma ^{-}+a^{(k)}\sigma ^{+}).\label{JC}
\end{equation}

Here, the first two terms correspond to the qubit and the resonator, with $a$ and $a^{\dag}$ being the annihilation and the creation operators of the resonators photon field. The third term corresponds to the  the multi-photon qubit-resonator interaction, where $\sigma^\pm$ are the qubit raising/lowering operators and $\hbar g_{(k)}$ is the coupling energy for $k-$photon processes.  

\begin{figure}[tph]
	\includegraphics[width=4cm]{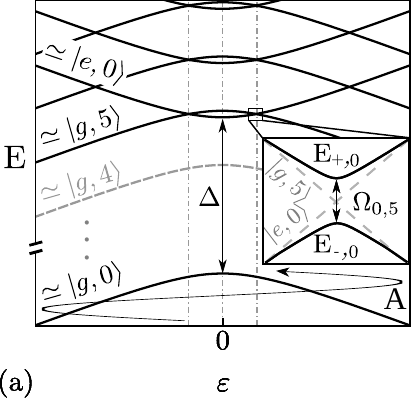}
	\includegraphics[width=4.5cm]{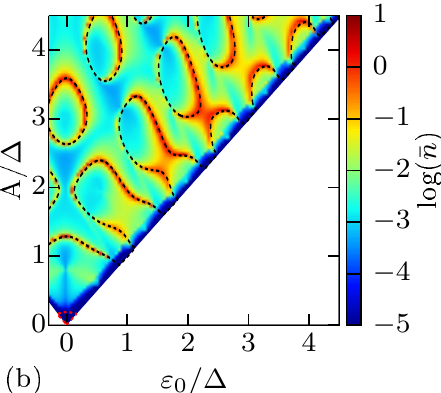}
	\caption{(color online). (a) Energy levels $E$ of a strongly coupled superconducting qubit-resonator system, as a function of the energy bias $\protect\varepsilon _{0}$ of the qubit. In the AIM model, the system mimics an array of beam splitters placed at avoided-level crossings. 
	(b) The average number of photons in the resonator in logarithmic scale, calculated by the AIM as a function of the qubit energy bias and the driving amplitude. The black dashed lines correspond to resonance condition given by Eq.~(\ref{condition}). The obtained  maximum number of photons in the resonator is approximately  8 for system parameters: $\protect\omega _{\mathrm{r}}/2\protect\pi =2.5$ GHz, $\protect\omega =3\protect\omega _{\mathrm{r}}$, $\Delta /h=12.2$ GHz and for simplicity $\Omega_{m, m+5}/h~=~10$~MHz.}
	\label{Fig:N_avg}
\end{figure}

The bare qubit-resonator system states are presented in the qubit-photon basis $|e/g, n\rangle$ and the corresponding eigenenergies of the system with $n$ photons are $E_{g,e} + \hbar\omega_r \left(n+1/2 \right)$, which can be seen by neglecting the interaction term in~(\ref{JC}). These energy levels are degenerated for a set of integer numbers $l$, $m$, where the multi-photon resonance condition $E_g + l \hbar \omega_r = E_e + m \hbar \omega_r$ is fulfilled. The energy levels of the $|e/g, n\rangle$ eigenstates for $\protect\omega _{r}/2\protect\pi =2.5$~GHz and $\Delta /h=12.2$~GHz are depicted in Fig.~\ref{Fig:N_avg}(a). For simplicity, in our numerical model, we consider only $k=5$ photon processes. 
The qubit-resonator interaction lifts the degeneracy for $l=m+k$, as is shown in the insert in Fig.~\ref{Fig:N_avg}(a). Therefore, close to resonance, the eigenstates of the system are dressed qubit-resonator states $E_{\pm,m}$ with energy level separation at avoided-level crossings $\Omega_{l, m}$.\cite{Vogel2006} However, far from the avoided-crossings, the energy levels are well approximated by the bare qubit-resonator states $|e/g, n\rangle$.

By driving the system, i.e. changing the qubit energy bias as $\varepsilon(t)=\varepsilon _{0}+A\sin \omega t$, LZ tunneling occurs both between $|g, n\rangle$ and $|e, n\rangle$ states at $\epsilon=0$ with level separation $\Delta$ and between the dressed states $E_{\pm,m}$ at resonances $\omega_q=(5+m)\omega_r$ with level separations $\Omega_{m, m+5}$. A similar system with three-photon quantum Rabi oscillations was recently studied in Ref.~\onlinecite{Garziano2015}. Numerically, the time evolution is simulated as a sequence of LZ transitions at these avoided-crossings and adiabatic evolution of the bare-qubit resonator states. During LZ transition between $|g, n\rangle$ and $|e, n\rangle$, only the state of the qubit changes (energy is transferred between the qubit and the driving field). Whereas, at LZ transitions between the dressed qubit-resonator states, the state of both the qubit and the resonator changes, since this process is equivalent to photons absorption/emission between the resonator and the qubit. 

%During LZ transitions between $|g,k+n\rangle$ and $|e,n\rangle$, states of both qubit and resonator change, since this process is equivalent to photons absorption/emission between resonator and qubit. 

Our model was limited to 40 levels ($m\leq 20$) and the simulation was initialized at the ground state of the system $|g, 0\rangle$.  The system state was averaged over a number of periods N~=~20~000 of the driving field, to estimate the average photon number in the resonator as a function of the driving amplitude $A$ and qubit bias $\varepsilon_0$. For the set of parameters obtained from the experiment on the conventional Josephson junction qubit, the AIM simulation shows LZSM interference patterns with high average photon number areas, shown in Fig.~\ref{Fig:N_avg}(b). At these areas, as the average photon number in the resonator is increased by the non-thermal occupancy of the higher resonator states, increased photon emission as well as increased transmission of the resonator (in the case of probing the transmission of the resonator $\protect\omega_{p}\approx\protect\omega_r$) is expected. These patterns agree perfectly with the position of resonances between the Rabi-like oscillations and the resonator mode given by Eq.~(\ref{condition}). To achieve a better  agreement between the numerical model and the experiment on the Josephson junction qubit, coupling to additional degrees of freedom is required, as the LZSM interference pattern is strongly influenced by coupling to a bath.\cite{Blattmann15} This can be carried out by the AIM with quantum jumps that occur randomly during the time evolution of the system and lead to fluctuations and dissipation.\cite{Molmer93} Such approach could lead to a better understanding of the off-resonant driving of a qubit with strong dissipation which is important for many fields, such as LZSM interferometry itself, laser science in semiconductors \cite{Hughes11}, quantum diffusion \cite{Sepehrinia15,Yang15} etc.

\section{Conclusion}

In conclusion, we measured the emission from a resonator coupled to a strongly driven qubit as well as the transmission of a weak probing signal through the resonator. The qubit experiences Rabi-like oscillations and when the frequency of these transitions matches the resonant frequency of the resonator, the number of photons in the cavity is increased or decreased. This is experimentally observed as either photon emission, or amplification and attenuation of the resonator normal mode signal,  which can be referred to as lasing and cooling, respectively. 
 The driven qubit is described in terms of the LZSM interference, 
 where the sequential non-resonant non-adiabatic transitions result, 
 due to the interference, in Rabi-like oscillations. 

\section{ACKNOWLEDGMENTS}

The research leading to these results has received funding from the European
Community Seventh Framework Programme (FP7/2007-2013) under Grant No. 270843
(iQIT) and APVV-DO7RP003211. This work was also supported by the Slovak Research and Development Agency under the contract APVV-0808-12, APVV-0088-12, and APVV 14-0605,  BMBF (UKR-2012-028), and the State Fund for Fundamental Research of Ukraine (F66/95-2016). E.I. acknowledges partial support from the Russian Ministry of Education and Science, within the framework of State Assignment 8.337.2014/K. S.N.S. thanks S.~Ashhab for useful discussions and P.N. acknowledge helpful discussions with S.~Kohler and R.~Blattmann.

\section*{APPENDIX A: Theory}\label{APPENDIX A}

In the main text, we presented several results for the description of the
driven two-level system by the adiabatic-impulse method (AIM); for more
details about this model, see Ref.~\onlinecite{Shevchenko10} and references
therein. In particular, in the fast-passage limit, this model gives correct
expressions for the multi-photon Rabi oscillations in the system, where the
correctness is confirmed by the agreement with the rotating-wave
approximation (RWA). This appears as a wonder, since the fast-passage limit
is, strictly speaking, beyond the region of originally assumed validity of
the AIM. Moreover, even in the opposite limit of slow
passage, similar, Rabi-like oscillations appear. In this Appendix, 
we present in more details how those results are derived.

\subsubsection{Multiphoton Rabi oscillations}

Here, we first recall the results obtained for the Rabi oscillations in the
RWA. First, the textbook example is the weakly-driven two-level system, with
$A\ll \Delta $, which is considered close to the resonance, where the frequency $%
\omega $ is near the characteristic frequency of the two-level system $%
\omega _{\mathrm{q}}=\Delta E/\hbar $. With such assumptions, the RWA
describes Rabi oscillations \cite{Coh-Tan} with the frequency%
\begin{eqnarray}
\Omega _{\mathrm{R}}^{2} &=&\Omega _{\mathrm{R0}}^{2}+\delta \omega
^{2},\;\;\delta \omega =\omega _{\mathrm{q}}-\omega ,  \label{Om_R} \\
\Omega _{\mathrm{R0}} &=&\frac{\Delta A}{2\hbar \Delta E}.  \notag
\end{eqnarray}

Another version of RWA can be developed when the small value is the
adiabaticity parameter $\Delta ^{2}/(A\cdot \hbar \omega )\ll 1$, see e.g.
in Refs.~\onlinecite{Garraway97, Ashhab07, Oliver05, Shevchenko14}. This condition means
that $\Delta $ is small - then $\Delta E\approx \left\vert \varepsilon
_{0}\right\vert $. On the other hand, the above condition means that the
driving is strong and that the avoided region is passed fast. In this
strong-driving fast-passage limit the system is resonantly excited at $%
k\omega \simeq \omega _{\mathrm{q}}$, which corresponds to the $k$-photon
transitions with the frequency%
\begin{eqnarray}
\Omega _{\mathrm{R}}^{2} &=&\Omega _{\mathrm{R0}}^{2}+\delta \omega
^{2},\;\;\delta \omega =\omega _{\mathrm{q}}-k\omega ,  \label{with_Bessel}
\\
\Omega _{\mathrm{R0}} &=&\Delta J_{k}\left( \frac{A}{\hbar \omega }\right) .
\notag
\end{eqnarray}%
From here, in particular at weak driving, $A/\hbar \omega \ll 1$, only the
transition with $k=1$ is relevant, and with $\hbar \omega \approx \left\vert
\varepsilon _{0}\right\vert \approx \Delta E$ and with the asymptote $%
J_{1}(x)\approx x/2$ we obtain Eq.~(\ref{Om_R}). In the opposite limit of
strong driving, $A/\hbar \omega \gg 1$, another asymptote of the Bessel
function is relevant: $J_{k}(x)\approx \sqrt{\frac{2}{\pi x}}\cos \left( x-%
\frac{\pi }{4}\left( 2k+1\right) \right) $. These known results, presented
in this subsection, are needed for further comparison with the results of
the AIM.

\subsubsection{Rabi oscillations in AIM}

This subsection and the next one are devoted to the results obtained in AIM\
in two limiting cases. To start with, we note that the AIM was analyzed in
many publications, of which a review can be found in Ref.~\onlinecite%
{Shevchenko10}. There, and also in Ref.~\onlinecite{Ashhab07}, the resonance
conditions, the width of the resonances, and the frequency of the resulting
oscillations were studied. The resonance condition is written down in Eq.~(%
\ref{qubit-resonance}).

The AIM allows for analytical solution for the frequency of the Rabi-like
oscillations. In particular, the AIM predicts slow oscillations of the
qubit's occupation probabilities (see Ref.~\onlinecite{Shevchenko10}). Here, the
time dependence is given by the factor $\sin ^{2}n\phi $, where $n$ is the
number of periods passed and $\phi $ is defined by$\ $%
\begin{equation}
\cos \phi =\left( 1-P_{\mathrm{LZ}}\right) \cos \zeta _{+}-P_{\mathrm{LZ}%
}\cos \zeta _{-},  \label{fi}
\end{equation}%
where $\zeta _{\pm }=\widetilde{\zeta }_{1}\pm \widetilde{\zeta }_{2}$. If
the frequency of these oscillations is smaller than the driving frequency,
we can identify the factor $\sin ^{2}n\phi $ with the one corresponding to
oscillations with frequency $\Omega $, which is $\sin ^{2}\frac{\Omega }{2}t$.
\cite{Garraway97} During one driving period, the integer $n$ changes by
unity and this corresponds to changing the time $t$ by one period $2\pi
/\omega $. With this, we obtain the relation for the coarse-grained
oscillations: $\Omega =\frac{\omega }{\pi }\left\vert \phi \right\vert $,
which together with Eq.~(\ref{fi}) results in the equation~(\ref{cos(fi)}).
In addition, the amplitude of the Rabi-like oscillations is maximal when the
resonance condition for the driven qubit is fulfilled \cite{Shevchenko10,
Silveri15}:
\begin{equation}
\left( 1-P_{\mathrm{LZ}}\right) \sin \zeta _{+}-P_{\mathrm{LZ}}\sin \zeta
_{-}=0.  \label{qubit-resonance}
\end{equation}

So, we have the formula for resonances, Eq.~(\ref{qubit-resonance}); these
resonances can have constructive or destructive character, and in the former
case, the slow oscillations with the frequency $\Omega \ll \omega $, as given by (1), can take place. This can be used for arbitrary parameters, which is illustrated in Fig.~\ref{Fig:En_lev}. However, for deeper understanding, it is worthwhile to consider several limiting cases.

Consider first the \textit{strong-driving fast-passage limit}, assuming $%
\delta \ll 1$, $1-P_{\mathrm{LZ}}\ll 1$, and $A\gg \varepsilon _{0}$. Then
one can obtain (see also in Ref.~\onlinecite{Shevchenko10}): $\widetilde{\varphi }%
_{\mathrm{S}}\approx -\pi /4$ and%
\begin{equation}
\zeta _{-}\approx \frac{\pi \varepsilon _{0}}{\hbar \omega },\;\;\zeta
_{+}\approx \frac{2A}{\hbar \omega }-\frac{\pi }{2}.
\end{equation}%
The approximated resonant condition (\ref{qubit-resonance}) gives $\zeta
_{-}\approx k\pi $, which corresponds to the $k$-photon resonance condition,
$\left\vert \varepsilon _{0}\right\vert $ $\approx k\hbar \omega $ with
positive integer $k$. This means that the resonances take place at $\omega
\approx \omega ^{(k)}=\left\vert \varepsilon _{0}\right\vert /\hbar k$.
Consider small deviations $\delta \omega $ from this value, $\omega =\omega
^{(k)}+\delta \omega /k$. Then after some trigonometric derivations, we
obtain the expression for $\phi $. Next we can calculate the upper diabatic
level occupation probability, defined by the equation from Refs.~\onlinecite%
{Shevchenko10, Silveri15, Garraway97}:%
\begin{equation}
P_{\mathrm{up}}(n)=\frac{2\cos ^{2}\widetilde{\zeta }_{2}}{\sin ^{2}\phi }%
\sin ^{2}n\phi \text{.}
\end{equation}%
Then for the upper diabatic level we obtain the occupation probability $P_{%
\mathrm{up}}(t)$:

\begin{eqnarray}
P_{\mathrm{up}}(t) &=&\overline{P}(1-\cos \Omega _{\mathrm{R}}t),\;
\label{Pup(t)} \\
\overline{P} &=&\frac{1}{2}\frac{\Omega _{\mathrm{R0}}^{2}}{\Omega _{\mathrm{%
R}}^{2}},\;\;\Omega _{\mathrm{R}}^{2}=\Omega _{\mathrm{R0}}^{2}+\delta
\omega ^{2},\;\;\delta \omega =\!k\omega -\frac{\left\vert \varepsilon
_{0}\right\vert }{\hbar },  \notag \\
\Omega _{\mathrm{R0}} &=&\Delta {\frac{2 \hbar \omega }{\pi  A}}%
\left\vert \cos \left( \frac{A}{\hbar \omega }-\frac{\pi }{4}(2k+1)\right)
\right\vert .  \notag
\end{eqnarray}%
Thus, in this limit, we obtain the multiphoton Rabi oscillations; these were
analyzed in detail in Refs.~\onlinecite{Ashhab07, Silveri13, Shevchenko14}. We
note that Eq.~(\ref{Pup(t)}) is in remarkable agreement with the
multi-photon Rabi oscillations described by RWA, Eq. (\ref{with_Bessel}). To
emphasize this accord, in this case we denoted $\Omega \equiv \Omega _{%
\mathrm{R}}$.

\subsubsection{Rabi-like oscillations}

Similarly to the above, one can consider other limiting cases. Here we
consider the limit of \textit{slow} and \textit{strong} driving with $\delta >1$, $P_{%
\mathrm{LZ}}\ll 1$, and $A\gg \Delta $, assuming in addition $\varepsilon
_{0}=0$. Then we obtain \cite{Shevchenko10}: $\widetilde{\varphi }_{\mathrm{S%
}}\approx -\pi /4$, and%
\begin{equation}
\zeta _{-}\approx 0,\;\;\zeta _{+}\approx \frac{2A}{\hbar \omega }-\pi .
\end{equation}%
The resonance condition (\ref{qubit-resonance}) gives $\frac{A}{\hbar \omega
}=\frac{\pi }{2}m$ with integer $m$. For odd and even $m$, the interference
bears constructive and destructive interference; this can be seen from the
expression for the adiabatic upper-level occupation probability \cite%
{Shevchenko10}:%
\begin{equation}
P_{+}(n)=\frac{P_{\mathrm{LZ}}}{\cos ^{2}\frac{A}{\hbar \omega }+P_{\mathrm{%
LZ}}\sin ^{2}\frac{A}{\hbar \omega }}\sin ^{2}n\phi \text{.}  \label{P+(n)}
\end{equation}%
The constructive interference for $m=2k+1$ results in the Rabi-like
oscillations; for illustration see Fig.~\ref{Fig:En_lev}.

Consider the frequency in the vicinity of the constructive resonance: $%
\omega =\omega ^{(k)}+\delta \omega /m$, where it is slightly shifted from $%
\omega ^{(k)}=2A/\pi \hbar m$. Developing in $\delta \omega $, we obtain
an expression for $\phi $ and then, from (\ref{P+(n)}) we get the
coarse-grained oscillations, described by the upper-level occupation
probability $P_{+}(t)$, its average value $\overline{P}$ and frequency $%
\Omega $:

\begin{eqnarray}
P_{+}(t) &=&\overline{P}(1-\cos \Omega t),\;\;\overline{P}=\frac{1}{2}\frac{%
\Omega _{0}^{2}}{\Omega ^{2}},\;\;\Omega ^{2}=\Omega _{0}^{2}+\delta \omega
^{2},  \notag \\
\Omega _{0} &=&\frac{2}{\pi }\sqrt{P_{\mathrm{LZ}}}\omega ,\;\;\delta \omega
=(2k+1)\omega -\frac{2A}{\pi \hbar }.  \label{P+(t)}
\end{eqnarray}%
This means that at $\delta \omega =0$, the oscillations are maximal with the
frequency defined by the LZ transition probability, $\Omega _{0}\propto
\sqrt{P_{\mathrm{LZ}}}$, which makes it much smaller than the driving
frequency, $\Omega _{0}\ll \omega $. These approximate formulas are
demonstrated in Fig.~\ref{Fig:En_lev}(b) to quantitatively well describe the
exact solution.

\begin{figure}[]
\includegraphics[width=7cm]{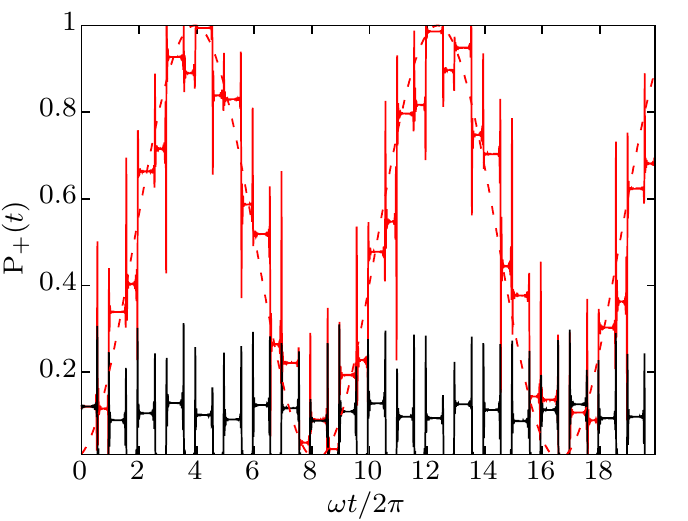}
\caption{(color online). The upper-level occupation probability $P_{+}(t)$
as a function of time for non-zero bias. Parameters are the following: $%
A/\Delta =15.75$, $\hbar \protect\omega /\Delta =0.05$, $\protect\varepsilon_{0}/\Delta=5.1$ and $5.17$ for the black and red curve respectively. }
\label{Fig:P+(t)_e0}
\end{figure}

We repeatedly emphasize that here, we started from the adiabatic picture with
small driving frequency $\omega \lesssim \Delta ^{2}/A\ll \Delta $ and with
a small probability of non-adiabatic transitions $P_{\mathrm{LZ}}\ll 1$, and
then in the AIM we obtained slow-frequency oscillations. These
oscillations were studied in Ref.~\onlinecite{Zhou14} both experimentally and
numerically. Here we note that they are described by a factor $\sin
^{2}n\phi \sim \sin ^{2}\frac{\Omega }{2}t$. Accordingly, in this picture
such oscillations can be termed as LZSM-Rabi or Rabi-like oscillations.

In addition, other limiting cases can be considered. As another interesting
situation, consider slow-passage strong-driving regime, similar to above,
but for non-zero bias $\varepsilon _{0}$. Namely, we assume $A\gg
\varepsilon _{0}\gg \Delta $, and obtain: $\widetilde{\varphi }_{\mathrm{S}%
}\approx -\pi /2$ and%
\begin{equation}
\zeta _{-}\approx \frac{\pi \varepsilon _{0}}{\hbar \omega },\;\;\zeta
_{+}\approx \frac{2A}{\hbar \omega }-\pi .
\end{equation}%
Then for the oscillations we obtain the frequency

\begin{eqnarray}
\Omega ^{2} &=&\Omega _{0}^{2}+\delta \omega ^{2},\;\;\delta \omega
=(2k+1)\omega -\frac{2A}{\pi \hbar }, \\
\Omega _{0} &=&\frac{2}{\pi }\sqrt{P_{\mathrm{LZ}}}\omega \left\vert \cos
\frac{\pi \varepsilon _{0}}{2\hbar \omega }\right\vert .  \notag
\end{eqnarray}%
Remarkably, admitting here $\varepsilon _{0}=0$, we obtain the correct
result, Eq.~(\ref{P+(t)}), even though we assumed here $\varepsilon _{0}\gg
\Delta $. Note also the strong dependence of the Rabi-like frequency $\Omega
_{0}$ on the bias $\varepsilon _{0}$. This is demonstrated in Fig.~\ref%
{Fig:P+(t)_e0}; compare this with Fig.~\ref{Fig:En_lev}(b) plotted for the
same parameters but zero bias, $\varepsilon _{0}=0$.

\section*{APPENDIX B: Experimental details}\label{APPENDIX B}

Two Josephson junction qubits were fabricated in the
 central part of a resonator by conventional shadow evaporation technique. The loop size of the qubits is $5\times 4.5~\mu $m$^{2}$ and each loop is
 interrupted by six Josephson junctions, of which the three smallest, sized $%
 0.2\times 0.3~\mu $m$^{2}$, $0.2\times 0.2~\mu $m$^{2}$ and $0.2\times
 0.3~\mu $m$^{2}$, determine the qubit dynamics. The additional Josephson
 junction provides coupling between the qubits as well as a qubit resonator
 coupling. By applying a certain energy bias, one of the qubits can be set to a
 localized state, while the second is in the vicinity of its degeneracy
 point. This way, we can measure the qubits separately. 
 
\begin{figure}[h!]
 	\includegraphics[width=7.0cm]{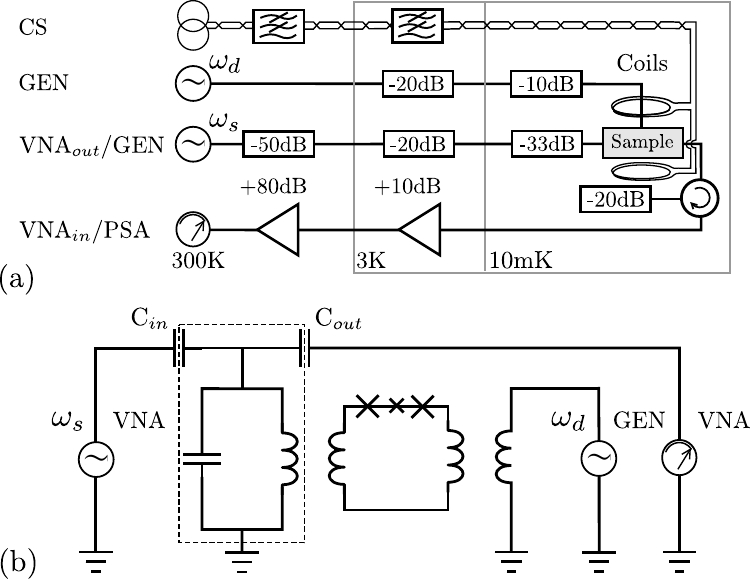}
 	\caption{(color online). The experimental setup scheme for transmission/emission
 		measurements carried out by a vector network analyser (VNA)/power spectrum analyzer (PSA).
 		During an emission measurement, to demonstrate the amplification presented in Fig.~\ref{Fig:Pavol_s}(c), an additional probe signal was applied from a generator (GEN) at $\omega_s$.
 		The qubit was biased by dc magnetic field of two superconducting coils mounted to a copper sample holder in Helmholtz geometry. The coils were fed by a dc current source (CS) filtered by carbon powder filters placed at 3K plate. The qubit was strongly driven by a microwave signal generator (GEN) at frequency $\omega /2\pi$ biasing an excitation loop through an additional coaxial line.   
 		(b) The simplified scheme of the measurement. The resonator's transmission was measured at $\protect\omega_s$ by vector network analyzer (VNA). The qubit was inductively coupled to the resonator and to a separate excitation loop as well, to drive the qubit at $\omega_d$.}
 	\label{Fig:MeasScheme}
 \end{figure} 
 
 The sample was thermally anchored to the mixing chamber of a dilution
 refrigerator, which maintained a temperature of about 30 mK during the experiment. 
 The scheme of our measurement set-up is shown in  Fig.~\ref{Fig:MeasScheme}(a).
 Both, the transmission and the spectral power density of the resonator at
 fundamental frequency $\omega _{\mathrm{r}}/2\pi$ were measured. The
 transmission/ emision of the resonator was measured by a vector network 
 analyser/ power spectrum analyzer. The input line was heavily filtered by a set of thermally
 anchored attenuators at 3K plate (20dB) and at the mixing chamber (-33dB).
 The output line was isolated by a cryogenic circulator which was placed
 between the sample and the self-made SiGe cryogenic amplifier mounted at
 the 3K-plate. The qubit, biased by an external dc magnetic field, was strongly
 driven by a microwave signal generator at frequency $\omega /2\pi $ through
 an additional coaxial line. The resonance frequency and the quality factor of the resonator's fundamental
 mode were determined from the transmission spectra of the coplanar waveguide resonator
 taken at weak probing. The simplified scheme of the measurement is shown in Fig.~\ref{Fig:MeasScheme}(b).     
The quantum phase slip qubit samples are fabricated using a process similar to Ref.~\onlinecite{Peltonen13}: First, a NbN film of thickness $d\approx 2-3$ nm is deposited on a Si substrate by dc reactive magnetron sputtering. Proceeding with the uniform NbN film, coplanar resonator groundplanes as well as the transmission lines for connecting to the external microwave measurement circuit are patterned in a first round of electron beam lithography (EBL) and subsequently metallized in an electron beam evaporator. In a second EBL step, the loops with constrictions as well as the resonator center line are patterned using a high resolution negative resist. Reactive ion etching (RIE) in CF$_4$ plasma is then used to transfer the pattern into the NbN film.

\bibliography{references}

\end{document}